\documentstyle[prb,aps,multicol,graphicx,psfrag]{revtex}

\def\Journal#1#2#3#4{{#1} {\bf #2}, #3 (#4)}
\def\Proc#1#2#3{{#1}, #2 (#3)}
\def\Site#1#2#3#4{{#1}/#2, #3 (#4)}

\def\CM{\em cond-mat}

\def\EPL{\em Europhys. Lett.}

\def\ICPS{\em Proc. 24$^{th}$ Int. Conf. on Phys. of Semicond.}

\def\JPC{\em J. Phys.: Condens. Matter}

\def\PLM{\em Phil. Mag.}
\def\PLA{{\em Phys. Lett.}  A}
\def\PRL{\em Phys. Rev. Lett.}
\def\PR{\em Phys. Rev.}
\def\PRB{{\em Phys. Rev.} B}

\def\ZPB{{\em Z. Phys.} B}

\begin{document}
\draft

\onecolumn

\title{Intense AC field driven superlattices with barrier width dimerization}

\author{P.~H.~Rivera$^1$, N.~Studart$^1$, and P.~A.~Schulz$^2$}

\address{$^1$ Departamento de F\'{\i}sica, Universidade Federal de S\~ao
Carlos \\ 13564--200, S\~ao Carlos, S\~ao Paulo, Brazil}

\address{$^2$ Instituto de F\'{\i}sica `Gleb Wataghin', Universidade
Estadual de Campinas \\ 13083-970 Campinas, S\~ ao Paulo, Brazil}

\date{\today}

\maketitle
\begin{abstract}
The evolution of two coupled mini-bands, generated by alternating barrier 
widths dimerization of a superlattice, driven by intense AC fields is 
investigated. The present model delivers a useful framework for the 
transition between the analytical high frequency regime and the extreme low 
frequency limit described by models based on Fukuyama's {\it et al.} 
(Phys. Rev. B {\bf 8}, 5579 (1973)) proposal.
\end{abstract}


\begin{multicols}{2} 

The behaviour of semiconductor superlattices (SL)s driven by intense
AC fields has been a subject of growing interest in the last few years
\cite{dunlap,holt92a,paulo,keay,jauho95}. The theoretical prediction of
isolated mini-band collapses\cite{holt92a}, the effect of the AC field
on multi-mini-bands and on disordered SLs
\cite{hone93,holt95a,holt95b,drese96} are among the interesting problems 
studied. Multi-mini-bands systems may be reduced to a simpler one, 
without loss of the main features, with
the suggestion of two almost symmetric mini-bands obtained by 
a dimerization process on a SL\cite{hone93}. 
This dimerized SL can be produced by an alternated sequence 
of, (1), two quantum wells, or, (2), two barriers, of different widths, 
keeping the barrier or quantum well (QW) 
widths constant for (1) and (2), respectively.

These two almost symmetric mini-bands may be strongly coupled by a AC 
field, whereas remaining isolated from other higher electronic mini-bands.
Initially suggested and discussed by Hone and Holthaus \cite{hone93}, 
 the effect of AC field on
dimerized SLs of alternated quantum wells has been 
systematically studied, either analytically 
in a high frequency limit\cite{zhao97} or
numerically for a wide range of field frequencies \cite{pablo98,pablo99}.

On the other hand, SLs
with alternated barriers have been studied for a limited 
parameter range\cite{zhao96,zhao96a,bao98}.  
More general models\cite{jauho95,drese96,holtphil96}, 
based on Fukuyama's {\it et al.} model\cite{fuku} combine characteristics of 
both dimerization procedures and show an involved behaviour respect 
to the field frequency with no clear distinction of 
different dynamic localization regimes. Our results suggest that these general
models should be carefully considered because of the assumptions respect the
SL coupling with the field. 

In the present work, we show that SLs with alternated barrier widths
show a more complex behaviour of the quasi-energy spectra as a function 
of field intensity compared to SLs with alternating well widths. Our results 
are based on a heuristic model with its validity  investigated
by comparison with previous results that take into account explicitly a SL
potential profile.
 
In general there are two different mini-band collapse regimes: 
isolated mini-band-like 
and dimer-like collapses. Also, we establish three effective
hopping parameter ranges for the two level systems that show the Stark shift 
evolution of the SL mini-bands. Due to the variety of parameters involved, 
only a partial picture of the competition between isolated mini-band 
behaviour and dynamic localization due to mini-band interactions may be 
sketched for dimerized SL with alternating barrier widths. 

We assume a two-mini-band model in the presence of a strong AC
electric field, within a tight-binding framework, emulating the dimerized
SL by alternated barriers\cite{hone93}.  A linear chain is considered 
for this SL, where each single ``atomic" $s$-like site orbital  
is associated to one quantized energy level of a QW.  The hopping 
parameters describe the coupling between the QW levels through the SL 
barriers.  The applied AC fields are parallel to the chain.  Hence, our 
model is described by the Hamiltonian $ H= H_o+H_{int}$, considering 
nearest neighbour interaction only:

$$ 
H_o=\sum_{\ell}E_{\ell}|\ell><\ell|+
$$
$$
{1\over 2}\sum_{\ell}\bigg\{\bigg [V_1 
{{(-1)^{\ell}+1}\over 2}+V_2 {{(-1)^{\ell +1}+1}\over 2}\bigg ]|\ell><\ell+1| 
$$
\begin{equation}
\bigg [V_1 {{(-1)^{\ell +1}+1}\over 2}+ 
V_2{{(-1)^{\ell}+1}\over 2} \bigg ]|\ell +1><\ell |\bigg \} \label{e1} 
\end{equation}
\begin{equation} 
H_{int}=eaF\cos \omega t\sum _{\ell}|\ell>\ell<\ell| \label{e2}
\end{equation}

\noindent 
where $E_{\ell}$ is the {\it s} orbital energy and $\ell $ is the
index site; $\omega$ and $F$ are the AC field frequency and amplitude,
respectively.  $a$ is the chain lattice parameter, $e$ is the 
electron charge. $V_1$
e $V_2$ are  hopping parameters that alternate along the chain. 
 The treatment of the
time-dependent problem is based on Floquet states $|\ell,m>$ where $m$ is
the photon index.  We follow the procedure first developed by Shirley
\cite{shirley} which consists in the transformation of the time-dependent
Hamiltonian into a time-independent infinite matrix which must be
truncated.  The matrix elements are:

$$
[({\mathcal E}-m\hbar\omega -E_{\ell})\delta_{\ell '\ell}-
$$
$$
\left[V_1{{(-1)^{\ell}+1}\over 2}+V_2 {{(-1)^{\ell +1}+1}\over 2}\right]\delta_{\ell',\ell-1}
$$
$$
\left[V_1 {{(-1)^{\ell +1}+1}\over 2}+V_2{{(-1)^{\ell}+1}\over 2}\right]\delta_{\ell',\ell+1}]\times
$$
\begin{equation}
\delta _{m'm}=F'\ell\delta_{\ell'\ell}(\delta _{m',m-1}+\delta _{m',m+1})
\label{e3} 
\end{equation}

\noindent 
where $F'={1\over2}eaF$.  The dimension of the matrix is
$L(2M+1)$, where $L$ is the number of ``atomic'' sites, while $M$ is the
maximum photon index.  We choose $M$ in order to satisfy a convergence
condition:  symmetric spectra relative to the Quasi Brillouin Zones (QBZs)
edges\cite{holt92b}.  The first QBZ is spanned in the range
$-\hbar\omega/2\leq {\mathcal E}\leq\hbar\omega/2$\ , where $\mathcal E$ 
represents the eigenvalues of the Floquet matrix, also called as quasi energy.

In what follows, we show the quasi-energy spectra as a function
of the electric field intensity, in units of potential drop in through a chain
lattice parameter, $eaF$. We consider AC external fields with wavelengths 
that are many times longer than the length of the SL and are linearly 
polarized along the SL growth direction. We choose to keep the frequency 
field constant ($\hbar\omega=0.5$ meV, corresponding to 
$\nu\simeq 0.1$ THz) along the present analysis, while ``tuning'' the SL 
parameters (varying the bare mini-band gap $E^0_g=2|V_1-V_2|$, while  
$E_{\ell}=0$ meV fixed for all $\ell$ sites). This procedure reveals to
be very useful for a clear identification of different dynamic 
localization regimes, for a wide frequency range ($E^0_g/\hbar\omega\ll 1$ to
$E^0_g/\hbar\omega\gg 1$), because 
of the wide possibilities for independent tight-binding parameters 
variation. Our calculations consider, in all cases, 
$L=12$ sites and $M=80$ photons. The hopping parameters are given
for each case, in the corresponding figure.

Depending on the tight-binding parameters, that determine the bare 
electronic structure, there are qualitatively different behaviours 
of the dressed states with increasing AC field intensity. 
First, a distinction is given by the bare mini-band width, 
$\Delta^0$, to field frequency ratio:
$\Delta^0/\hbar\omega<1$ and $\Delta^0/\hbar\omega>1$ situations 
are illustrated in parts (a) and (b), respectively, for figures (2-6). 
A second important feature is the mini-band gap to frequency ratio and 
the figures cover examples across the whole range from 
$E^0_g/\hbar\omega\ll 1$ to $E^0_g/\hbar\omega\gg 1$. 
We also will see that the Stark shift drives to resonances between 
mini-band replicas, leading to an interesting interplay between 
inter-mini-band and intra-mini-band couplings due to the AC field for
$\Delta^0/\hbar\omega>1$. Indeed, in this situation and for
low field intensities intra-mini-band coupling is dominant 
leading to strong anti-crossings at the QBZ edges, as can be observed in 
figures (2-6). For high field intensities one could look for effective 
two level systems in order to describe the Stark shift. 
One of the main questions to be answered concerns the identification and 
evolution of isolated mini-band behaviour in SL with barrier width 
dimerization, having in mind the well width dimerized 
SLs\cite{pablo98,pablo99}, for which no isolated mini-band behaviour 
occurs for field intensities above the dynamic breakdown.

\begin{figure*}[h]
\begin{center}
\includegraphics[scale=0.41]{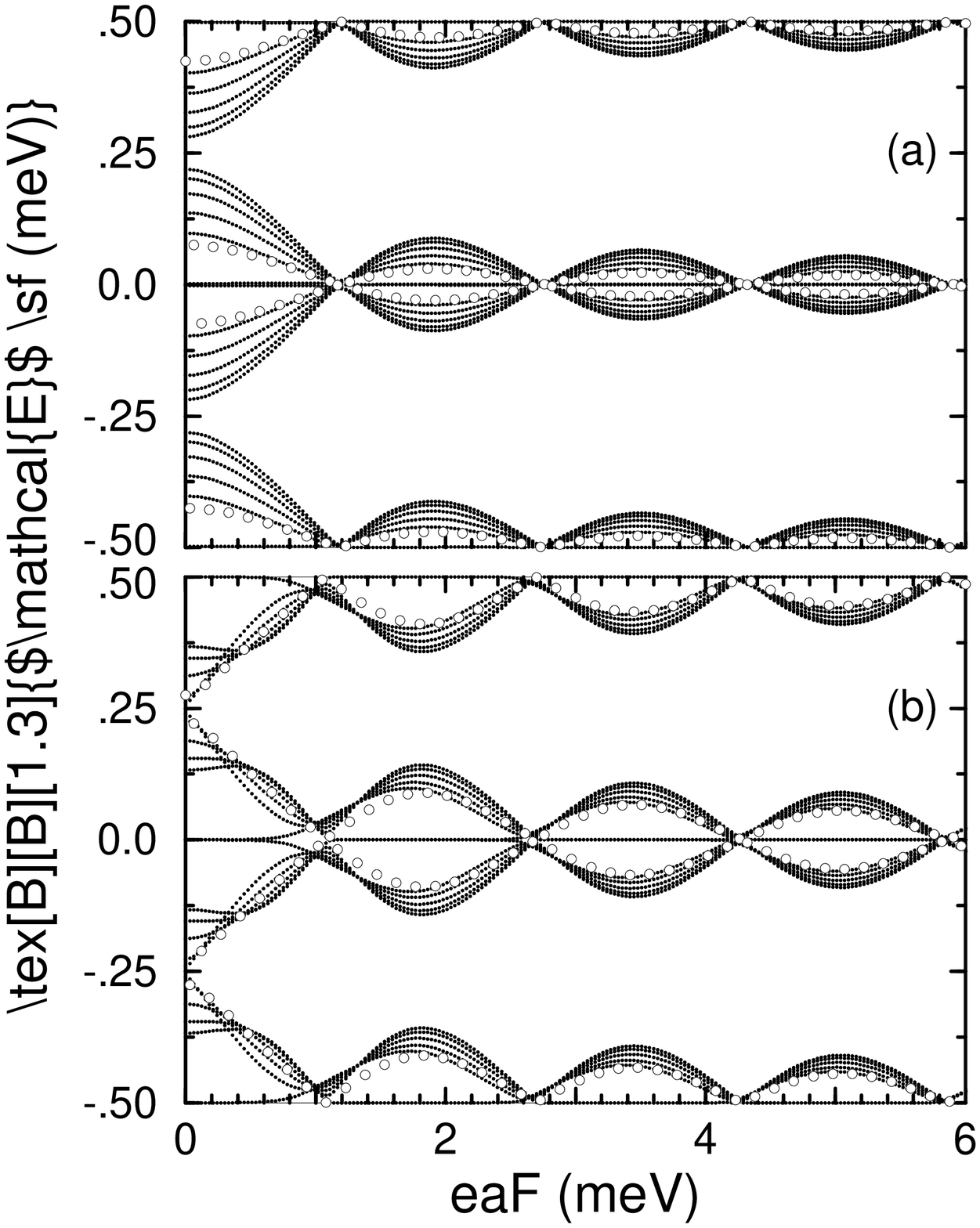}
\begin{minipage}{8cm}
\caption{Quasi energy spectra for dimerized SLs (dots) and dimers (white dots)
for $\Delta^0/\hbar\omega<1$ in both cases:
(a) $V_1=0.15$, $V_2=0.075$, $\Delta^0=0.15$, $E_g^0=0.15$ meV and 
$E_g^0/\hbar\omega=0.3$, $V_{\mbox{eff}}=0.075$ meV.
(b) $V_1=0.30$ e $V_2=0.075$, $\Delta^0=0.15$, $E_g^0=0.45$ meV and
$E_g^0/\hbar\omega=0.9$, $V_{\mbox{eff}}=0.225$ meV.
\label{f1}}
\end{minipage}
\end{center}
\end{figure*}

In the high frequency regime, $E^0_g/\hbar\omega\ll 1$ and 
in $\Delta^0/\hbar\omega<1$ situation, the dressed 
mini-bands resembles the results of an
analytical solution given by Bao\cite{bao98}, reformulated as:

\begin{equation}
{\mathcal E}=\pm\sqrt{(V_1^2+V_2^2)-2V_1V_2\cos(k_za)}\bigg |J_o\bigg(
{eaF\over\hbar\omega}\bigg)\bigg| \label{e4} 
\end{equation}

\noindent
where the mini-band gap, as a function of the field intensity, is defined by 
$E_g=2|V_1-V_2|J_0(eaF/\hbar\omega)$. The mini-band width is given by 
$\Delta=2V_1J_0(eaF/\hbar\omega)$, for $|V_2|>|V_1|$; and 
$\Delta=2V_2J_0(eaF/\hbar\omega)$, for $|V_1|>|V_2|$. On the other hand,
in $\Delta^0/\hbar\omega>1$ situation, the coupling of the dressed 
mini-bands is strong for low field intensities and this is not
predicted by Eq.(\ref{e4}); only for high field intensities, the evolution 
of dressed mini-bands is expressed by this.
For the regime, $E^0_g/\hbar\omega<1$ and in $\Delta^0/\hbar\omega<1$ 
situation, the behaviour based in
Eq.(\ref{e4}) also is observed, as can be 
clearly identified in the numerical results shown in Fig.(\ref{f1}). 

A further remark concerns to dependence of the surface states relative to 
the hopping parameters ratio. For $|V_2|>|V_1|$, the surface states merge 
into the mini-bands, while for $|V_1|>|V_2|$ two ``deep'' surface states 
appear, independently from the field intensity, as show in Fig.\ref{f1}.

In this regime, and for $E^0_g/\hbar\omega\ll 1$, 
no signatures of isolated 
mini-bands are identified; but it should be noticed that a mini-band collapse 
mechanism is given by the mini-band width modulation according to 
$J_0(eaF/\hbar\omega)$, associated to the mini-gap collapses, $E_g=0\ .$ 
For larger $E^0_g$, isolated mini-band like collapses start to be identified, 
while the dynamic localization mechanism of interacting mini-band replicas 
can be described by two level systems with an 
{\it effective hopping parameter}, $V_{\mbox{eff}}$. We will call such 
two level system as an {\it effective dimer}. The spectra of the 
{\it effective dimer} is also shown 
in Fig.1 (open circles). Here $V_{\mbox{eff}}=|V_2-V_1|$, as expected 
from the analytical result for $E_g$, derived from Eq.(4). 

\begin{figure*}[h]
\begin{center}
\includegraphics[scale=0.41]{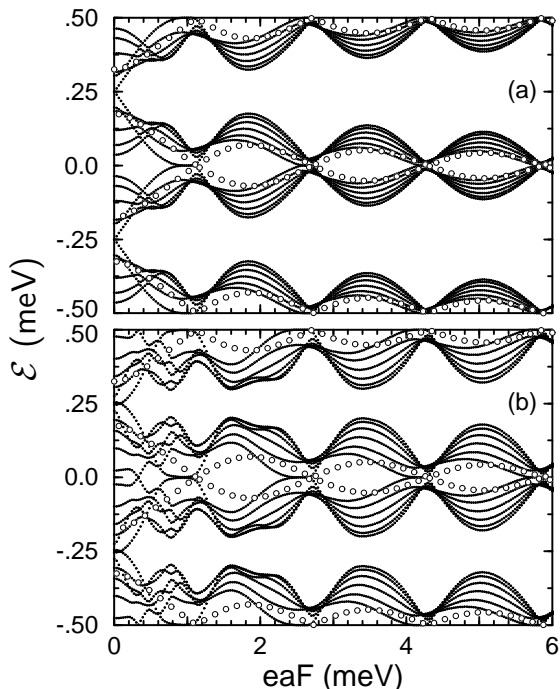}
\begin{minipage}{8cm}
\caption{Quasi energy spectra for dimerized SLs (dots) and dimers (white dots)
(a) $V_1=0.15$, $V_2=0.325$, $\Delta^0=0.30$, $E_g^0=0.35$ meV and 
$E_g^0/\hbar\omega=0.7$, $V_{\mbox{eff}}=0.175$ meV.
(b) $V_1=0.30$ e $V_2=0.475$, $\Delta^0=0.60$, $E_g^0=0.35$ meV and
$E_g^0/\hbar\omega=0.7$, $V_{\mbox{eff}}=0.175$ meV.
\label{f2}}
\end{minipage}
\end{center}
\end{figure*}

Fig.2 show quasi energy spectra for SLs with $E^0_g$ and $\Delta^0$
of the same order of those in Fig.1. Nevertheless, the 
spectra are quite different, markedly at low field intensities. The 
important difference is related to the role of the surface states, merged 
in the mini-bands in the spectra of Fig.2. The Stark shift of the mini-bands, 
however, is still well described by the same class of {\it effective dimer} 
for high field intensities. Although derived from a high frequency limit, 
the {\it effective dimer} describes the Stark shift and consequent crossing 
and anti-crossings of the mini-bands for the relatively wide 
$E_g^0/\hbar\omega\le 1$ range.

\begin{figure*}[h]
\begin{center}
\includegraphics[scale=0.41]{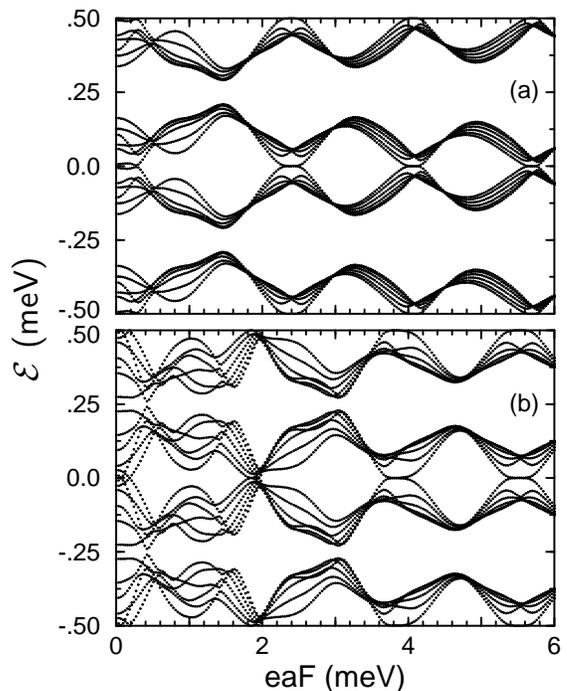}
\begin{minipage}{8cm}
\caption{Quasi energy spectra for dimerized SLs (dots) and dimers (white dots)
(a) $V_1=0.15$, $V_2=0.525$, $\Delta^0=0.30$, $E_g^0=0.75$ meV and 
$E_g^0/\hbar\omega=1.5$, $V_{\mbox{eff}}=|V_2-V_1|\longrightarrow V_2$.
(b) $V_1=0.30$ e $V_2=0.75$, $\Delta^0=0.60$, $E_g^0=0.90$ meV and
$E_g^0/\hbar\omega=1.8$, $V_{\mbox{eff}}=|V_2-V_1|\longrightarrow V_2$.
\label{f3}}
\end{minipage}
\end{center}
\end{figure*}

For lower frequencies, i.e., $E_g^0/\hbar\omega \sim 1$, the modulation 
of the mini-band width and the mini-band gap do not resemble anymore the 
simple analytical model of Eq.(4). In fact, dimer-like dynamic 
localization occurs for $1<E_g^0/\hbar\omega<2$, irrespective to the  
$\Delta^0/\hbar\omega$ ratio. But, the Stark shift can not be described by an 
{\it effective dimer} for a wide field intensity range. The resonances of
mini-band replicas induced by Stark shift may be fitted with an 
{\it effective hopping parameter}, which evolves from 
$|V_2-V_1|$ to $V_2$.
An example of this evolution is show in Fig.\ref{f3}, where  spectra of 
corresponding two level systems are not shown. Here, quasi energy spectra 
for SLs with $1<E_g^0/\hbar\omega<2$ and $\Delta^0/\hbar\omega<1$, Fig.3.a, 
and $\Delta^0/\hbar\omega>1$, Fig.3.b, are depicted. For this frequency 
range signatures of isolated-like mini-band collapses are already seen, 
although such behaviour becomes clearer for even lower frequencies as will 
be seen below. 

Fig.3.a delivers a further important information concerning 
the validity of our model, when compared to Fig.3 from reference \cite{holt94},
where the energy gap at zero field is 1.7 meV and the photon energy 1 meV:
both spectra are similar and belongs to the same parameter range. 
This comparison confirms that our heuristic model describes very well the 
main features of the spectra of a dimerized SL driven by intense AC fields.

The description of the mini-bands Stark shift with the consequent 
induced crossings and anti-crossing by an {\it effective dimer} becomes 
feasible again for even lower frequencies, i.e., 
$2\le E_g^0/\hbar\omega$, but now the effective dimer is given 
by $V_{\mbox{eff}} = V_2$. Two examples of this situation are illustrated in 
Fig.4, where the multi-photon resonances are well
described by resonances in the effective dimer spectra.
Concerning the comparison of SLs to the corresponding two level systems, the 
dimer-like collapses of these mini-bands are observed and associated to 
the resonances of dimers with {\it effective hooping parameters} given by: 

\begin{equation}
V_{\mbox{eff}}=\left\{\begin{array}{ccc}
|V_2-V_1| & \mbox{for} & E_g^0/\hbar\omega\le 1 \cr
|V_2-V_1|\longrightarrow V_2 & \mbox{for} & 1<E_g^0/\hbar\omega<2 \cr
V_2 & \mbox{for} & 2\le E_g^0/\hbar\omega \cr
\end{array} \right.
\end{equation} 

The isolated mini-band behaviour with the associated dynamic localization for 
field intensities corresponding to zeros of $J_0(e2aF/\hbar\omega)$ can be 
clearly identified in the spectrum shown in Fig.4(a). The mini-band collapse 
for $eaF \approx 0.6$ meV corresponds to the fist root of the zero order 
Bessel function taking into account the period of the SL, $2a$. Signatures
of further isolated-like mini-band collapses for 
higher field intensities are barely identified in this 
case. However, increasing further $E^0_g/\hbar\omega$, clear isolated 
mini-band behaviour may be seen up to very high field intensities, 
irrespective to the presence of mini-band resonances, as can be seen 
in Figs.\ref{f5} and \ref{f6}.  

\begin{figure*}[h]
\begin{center}
\includegraphics[scale=0.4]{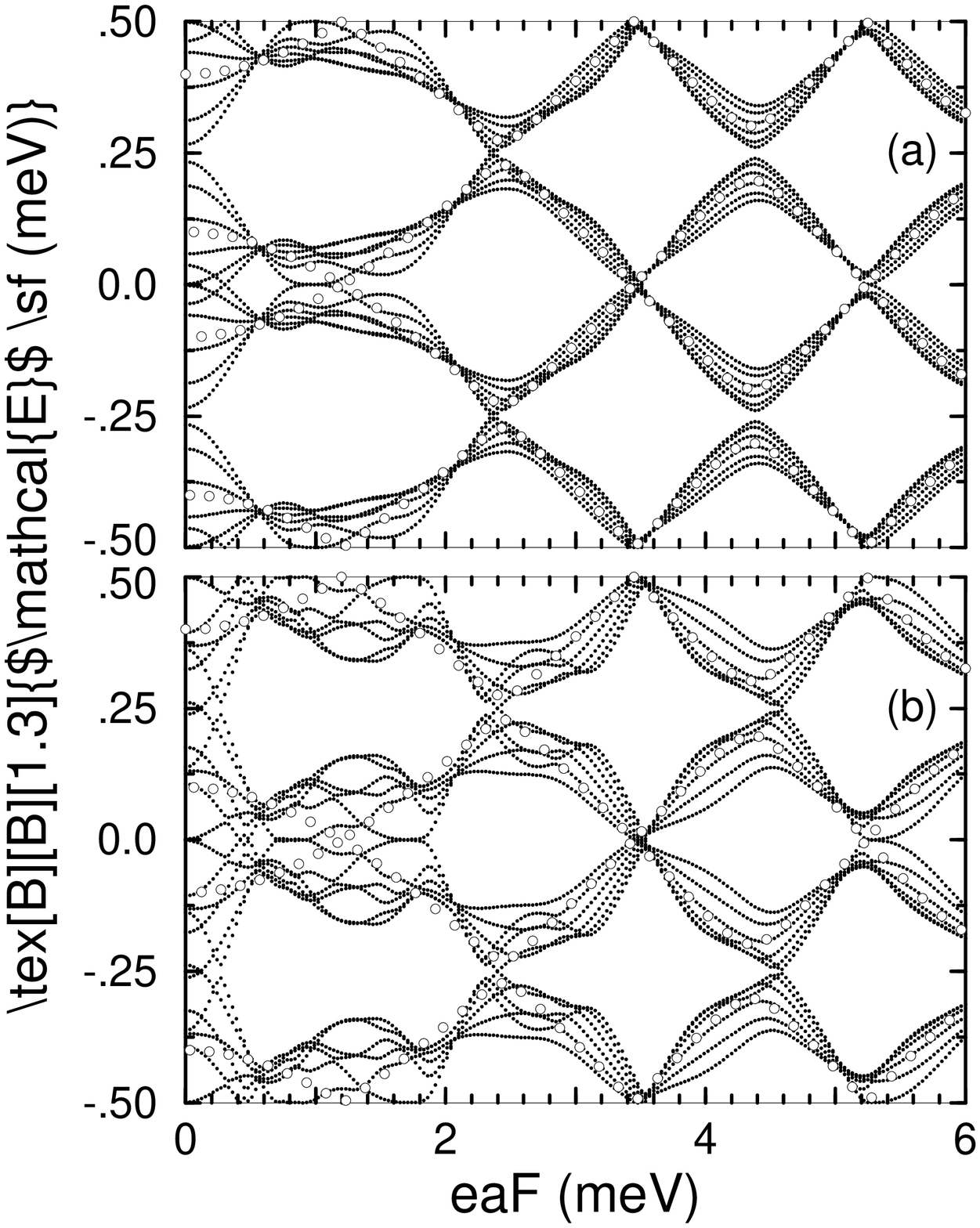}
\begin{minipage}{8cm}
\caption{Quasi energy spectra for dimerized SLs (dots) and dimers (white dots)
(a) $V_1=0.15$, $V_2=0.90$, $\Delta^0=0.30$, $E_g^0=1.50$ meV and 
$E_g^0/\hbar\omega=3.0$, $V_{\mbox{eff}}=0.90$ meV.
(b) $V_1=0.30$ e $V_2=0.90$, $\Delta^0=0.60$, $E_g^0=1.20$ meV and
$E_g^0/\hbar\omega=2.40$, $V_{\mbox{eff}}=0.90$ meV.
\label{f4}}
\end{minipage}
\end{center}
\end{figure*}

\begin{figure*}[h]
\begin{center}
\includegraphics[scale=0.41]{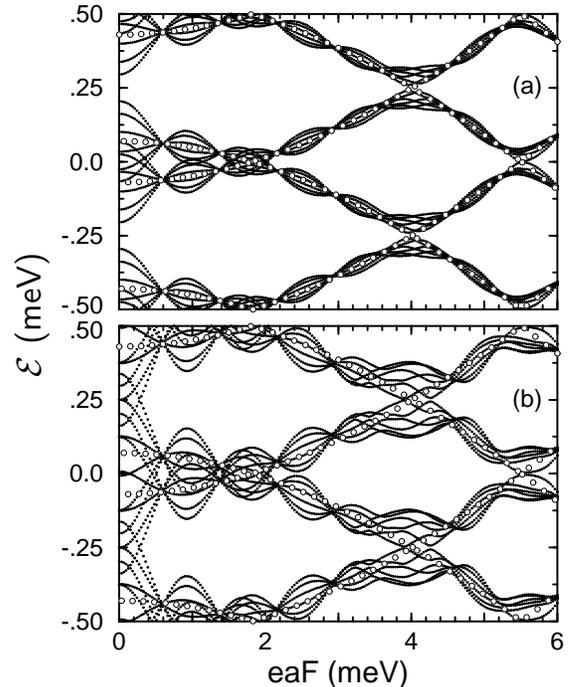}
\begin{minipage}{8cm}
\caption{Quasi energy spectra for dimerized SLs (dots) and dimers (white dots)
(a) $V_1=0.15$, $V_2=2.93$, $\Delta^0=0.30$, $E_g^0=5.56$ meV and 
$E_g^0/\hbar\omega=11.12$, $V_{\mbox{eff}}=2.93$ meV.
(b) $V_1=0.30$ e $V_2=2.93$, $\Delta^0=0.60$, $E_g^0=5.26$ meV and
$E_g^0/\hbar\omega=10.52$, $V_{\mbox{eff}}=2.93$ meV.
\label{f5}}
\end{minipage}
\end{center}
\end{figure*}

In the low frequency range, together with a high $V_2/V_1$ ratio, 
depicted in Figs.5 and 6, one has the limit of the Stark shift 
and related mini-band resonances well described by a system of two levels 
with an interaction described by the hopping parameter $V_2$. 
Less intense inter-dimer interaction, essentially described by the 
hopping parameter $V_1$, leads to a relatively small mini-band dispersion. 
In this limit the mini-band dispersion is not affected by mini-band 
interactions except for field intensities very close to multi-photon 
resonances. Nevertheless, strong mini-band interaction can be seen 
for $eaF \approx 4$ meV for a relatively wide mini-band case, Fig.5. 
Important, however, is that the isolated mini-band behaviour is recovered 
by further increasing the field intensity up to values for which a new 
multi-photon resonance occur at $eaF \approx 5.5$ meV. Such recovering of 
isolated mini-band behaviour for field intensities beyond a strong mini-band 
interaction does not occur with SL dimerized by alternating well 
widths \cite{pablo99}. The main difference lies on the fact that in the 
alternating well width case $E^0_g$ is mainly determined by the quantum 
well states. For field intensities that exceed the quantum well levels 
separation in energy a not recoverable breakdown occurs. In the present 
case of alternating barrier width, the bare mini-gap is given by
the difference between hopping parameters and the mini-band dispersion 
determined by the smaller one. Hence, the mini-band dispersion may be seen 
as a perturbation of the effective dimer and the breakdown is recoverable 
with further increasing of the field intensity.
 
For $E^0_g/\hbar\omega\gg 1$, the interaction between mini-bands is 
negligible for the field intensities shown in Fig.6. Actually, the avoided 
crossings are not resolved in the spectra, because the associated multi-photon 
resonances are for a very high photon number. Therefore, no dynamic breakdown 
is observed and the isolated mini-band character is preserved beyond such 
mini-band interactions.

\begin{figure*}[h]
\begin{center}
\includegraphics[scale=0.41]{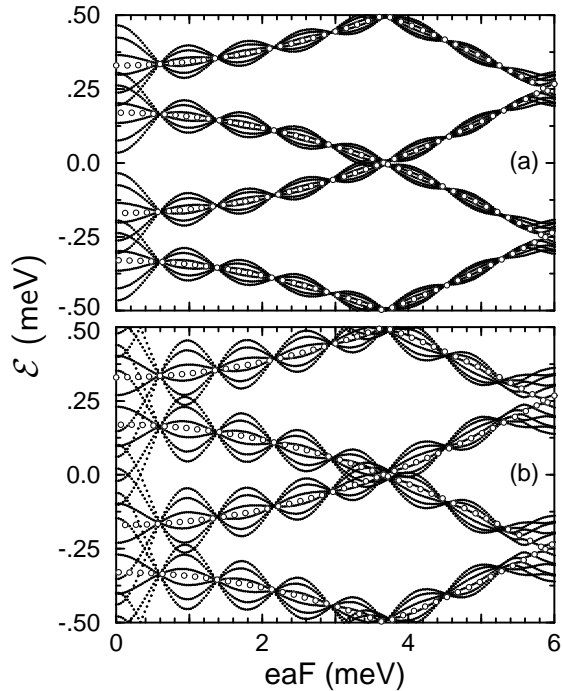}
\begin{minipage}{8cm}
\caption{Quasi energy spectra for dimerized SLs (dots) and dimers (white dots)
(a) $V_1=0.15$, $V_2=4.83$, $\Delta^0=0.30$, $E_g^0=9.36$ meV and 
$E_g^0/\hbar\omega=18.72$, $V_{\mbox{eff}}=0.90$ meV.
(b) $V_1=0.30$ e $V_2=4.83$, $\Delta^0=0.60$, $E_g^0=9.06$ meV and
$E_g^0/\hbar\omega=18.12$, $V_{\mbox{eff}}=0.9$ meV.
\label{f6}}
\end{minipage}
\end{center}
\end{figure*}

In conclusion, we propose a tight-binding model for SLs dimerized 
by alternating barrier widths. Our model, although heuristic, takes into 
account explicitly the barrier widths alternation in the proper hopping 
parameters. The heuristic character is only related to the construction of 
the bare SL. Within this framework, no further assumptions on the coupling 
of the SL with the AC field are made. The numerical results show that the 
model is appropriate irrespective to the $E^0_g/\hbar\omega$ ratio. One 
of the main findings of the present work is the inspection of the validity 
range of the analytical high frequency limit, as well of the models  
based on Fukuyama {\it et al.}\cite{fuku}, appropriate only for 
$E^0_g/\hbar\omega\gg 1$ and do not reproduce the high frequency limit. 
Furthermore, the present work elucidates the important differences between 
the dimerization by alternating barrier widths and by alternating well widths
\cite{pablo99}. For dimerization by alternating well widths, a clear breakdown 
is identified, while for the situation illustrated in Fig.6, alternating very
thin with relatively thick barriers, the isolated mini-band behaviour may be
periodically recovered as a function of increasing field intensity.
  
The authors acknowledge CAPES, FAPESP and CNPq for financial support.

\end{multicols}


\begin{references}

\bibitem{dunlap} D. H. Dunlap and V. M. Kenkre, \Journal{\PRB}{34}{3625} 
{1986}

\bibitem{holt92a} 
M. Holthaus, \Journal{\PRL}{69}{351}{1992}

\bibitem{paulo} 
P. S. S. Guimar{\~a}es, B. J. Keay, J. P. Kaminski, S. J. Allen, P. F. Hopkins,
A. C. Gossard, L. T. Florez, and J. P. Harbison, \Journal{\PRL}{70}{3792}{1993}
 
\bibitem{keay} B.J.Keay, S. Zeuner, S. J. Allen, K. D. Maranowski,
A. C. Gossard, U. Bhattacharya, and M. J. W. Rodwell, \Journal{\PRL}{75}
{4102}{1995}

\bibitem{jauho95} 
Jon Rotvig, Antti-Peka Jauho, and Henrik Smith, \Journal{\PRL}{74}{1831}{1995}

\bibitem{hone93}
Daniel W. Hone and Martin Holthaus, \Journal{\PRB}{48}{15123}{1993}

\bibitem{holt95a}
M. Holthaus, G. H. Ristow, and D. W. Hone, \Journal{\EPL}{32}{241}{1995}

\bibitem{holt95b}
Martin Holthaus and Gerard H. Ristow, \Journal{\PRL}{75}{3914}{1995}

\bibitem{drese96}
Klaus Drese and Martin Holthaus, \Journal{\JPC}{8}{1193}{1996}

\bibitem{holtphil96}
M. Holthaus and D. Hone, \Journal{\PLM}{74}{105}{1996}

\bibitem{zhao97}
Xian-Geng Zhao, \Journal{\JPC}{9}{L385}{1997}

\bibitem{pablo98}
P. H. Rivera and P. A. Schulz, \Proc{\ICPS}{Jerusalem, Israel; Ed. David
        Gershoni, World Scientific, Singapore}{1998}

\bibitem{pablo99}
P. H. Rivera and P. A. Schulz, \Site{\CM}{9908031}{Aug}{1999}

\bibitem{zhao96}
Xian-Geng Zhao e Qian Niu; \Journal{\PLA}{222}{435}{1996}

\bibitem{zhao96a}
Xian-Geng Zhao, G. A. Georgakis e Qian Niu; \Journal{\PRB}{54}{R5235}{1996}

\bibitem{bao98}
Shu-Qing Bao, Xian-Geng Zhao, Xin-Wei Zhang, and Wei-Xian Yan,
\Journal{\PLA}{240}{185}{1998}

\bibitem{shirley}
Jon H. Shirley, \Journal{\PR}{138}{B979}{1965}

\bibitem{holt92b}
M. Holthaus, \Journal{\ZPB}{89}{251}{1992}

\bibitem{fuku}
H. Fukuyama, R. A. Bari, and H. C. Fogedby, \Journal{\PRB}{8}{5579}{1973}

\bibitem{holt94}
Martin Holthaus and Daniel W. Hone, \Journal{\PRB}{49}{16605}{1994}

\end{references}
\end{document}